\documentclass[10pt]{iopart}
\expandafter\let\csname equation*\endcsname\relax
\expandafter\let\csname endequation*\endcsname\relax 
\usepackage{graphicx}
\usepackage{color}
\usepackage{cite}
\usepackage{amsmath}
\usepackage{iopams}
\usepackage{multirow}
\usepackage{makecell}
\usepackage[table,xcdraw]{xcolor}
\usepackage[hidelinks]{hyperref}
\hypersetup{
    colorlinks,
    linkcolor={blue!80!black},
    citecolor={blue!80!black},
    urlcolor={blue!80!black}}

\newcommand{\ket}[1]{\mbox{$\left|#1\right\rangle$}}
\newcommand{\bra}[1]{\mbox{$\left\langle#1\right|$}}
\newcommand{\braket}[2]{\mbox{$\langle#1|#2\rangle$}}
\newcommand{\set}[1]{\{#1\}}
\newcommand{\outerp}[2]{\ket{#1}\bra{#2}}

\newcommand{\ua}{\uparrow}
\newcommand{\da}{\downarrow}
\newcolumntype{?}{!{\vrule width 2pt}}
\newcommand{\ALS}[2]{#2}

\begin{document}

\title{Non-Markovianity and bound states in quantum walks with a phase impurity}

\author{B. Danac\i$^1$, G. Karpat$^2$, \.{I}. Yal\c{c}{\i}nkaya$^3$ and
A. L. Suba{\c s}{\i}$^1$
}
\address{$^1$ Department of Physics, Faculty of Science and Letters, Istanbul Technical University, 34469 Maslak, \.{I}stanbul, Turkey}
\address{$^2$ Faculty of Arts and Sciences, Department of Physics, Izmir University of Economics, \.{I}zmir, 35330, Turkey}
\address{$^3$ Department of Physics, Faculty of Nuclear Sciences and Physical Engineering, Czech Technical University in Prague, B\v{r}ehov\'a 7, 115 19 Praha 1-Star\'e M\v{e}sto, Czech Republic}
\ead{danacib@itu.edu.tr}

\begin{abstract}

We study the discrete-time quantum walk on the line with a single phase impurity. The spread and localisation properties of discrete-time walks initialized at the impurity site are affected by the appearance of bound states and their reflection symmetry. Here, we measure localisation by means of an effective localisation length and an effective participation ratio, which are obtained by averaging over all eigenstates and over all initial states, respectively. We observe that the reduced coin system dynamics undergoes oscillations in the long-time limit with the frequencies determined by the sublattice operator and the bound state quasi-energy differences. The oscillations give rise to non-Markovian evolution, which we quantify using the trace distance and entanglement based measures of non-Markovianity. Indeed, we reveal that the degree of the non-Markovian behaviour is closely related to the emergence of bound states due to the phase impurity. We also show that the considered measures give qualitatively different results depending on the number and symmetries of supported bound states. Finally, comparing localisation and non-Markovianity measures, we demonstrate that the degree of non-Markovianity becomes maximum when the walker is most localised in position space.

\end{abstract}

\pacs{
03.65.Yz, 
03.67.-a, 
05.40.Fb  
}

\vspace{2pc}
\noindent{\it Keywords}: quantum walk, bound states, localisation, non-Markovianity

\submitto{jpa}


\section{Introduction}

The theory of open quantum systems provides the necessary means to study and characterize the dynamics of quantum systems that are in an inevitable interaction with their surrounding environments~\cite{openbook}. It is well known that although closed systems evolve in time unitarily, dynamics of open quantum systems is no longer unitary due to the coupling to their environments. Such an interaction between the principal open system and the environment typically results in decoherence of the principal system, resulting in the loss of characteristic quantum properties such as coherent phase relations. From the standpoint of dynamical memory effects, time evolution of open quantum systems can be classified as exhibiting Markovian (memoryless) or non-Markovian behaviour. As a consequence of the increasing experimental control over quantum systems and the development of reservoir engineering techniques in recent years, the study of quantum non-Markovianity has become a significant line of research~\cite{rivas14,breuer16}. Various different methods have been introduced for quantifying and characterizing the non-Markovian behaviour in the dynamics of open quantum systems. Among others, the approaches based on the information dynamics between the open system and its environment have become prominent~\cite{fanchini14,lu10,luo12,rivas10,breuer09} since when information flows from the environment back to the system throughout the dynamics, the future states of the open system might depend on its earlier states.

Quantum walks have been proposed as the quantum counterpart of the classical random walks~\cite{aharonov1993} and they attracted considerable attention in the first place due to their quadratically faster spreading rates compared to their classical analogues~\cite{kempe2003}. In nearly two decades, the subject has gained its place in quantum computation as a substantial field of research on both the theoretical and the experimental sides~\cite{venegas-andraca2012}. It turned out, for example, that quantum walks are promising resources for developing new quantum algorithms~\cite{ambainis2003} and are universal for quantum computation~\cite{lovett2010}. They also provide a powerful framework for simulating physical systems~\cite{buerschaper2004,oka2005}, quantum state transfer \cite{kurzynski2011, zhan2014, yalcinkaya2015, stefanak2016, stefanak2017} and for examining the topological quantum matter~\cite{kitagawa2010,kitagawa2012,Cedzich2016,Cedzich2018a,Cedzich2018}. The open system dynamics of quantum walks on the other hand has also been extensively studied in existence of various noise channels~\cite{brun2003, romanelli2005, kendon2007, annabestani2010}. One of the main consequences of these studies is that decoherence may supress the spreading rate and give rise to a quantum to classical transition in the probability distribution of the walker. Some specific types of noise may even lead to Anderson localisation~\cite{schreiber2011}. Many of these theoretical works are supported by experiments performed over various physical setups as well, including ultracold atoms in optical lattices~\cite{karski2009}, trapped ions~\cite{schmitz2009, zahringer2010}, photons in a fibre loop~\cite{schreiber2010} and waveguide lattices~\cite{perets2008}.

\ALS{}{The connection between localisation and non-Markovianity has been studied in different settings, such as atomic impurities embedded in a disordered coupled cavity array ~\cite{lorenzo2017quantum}, in a quasi periodic Fermi lattice\cite{cosco2018memory} and in a Bose Hubbard lattice\cite{cosco2018bose}. These studies show that non-Markovian memory effects emerge  as excitations localise in the vicinity of the impurity.} 
In quantum walks, non-Markovian effects can be analysed without considering an extra external environment and focusing on the coin space reduced dynamics where the spatial degree of freedom is traced out \cite{hinarejos2014}. In that case, the position space itself is treated as the environment of the coin space. Even though this reduced dynamics is known to be non-Markovian for the standard quantum walk, the presence of decoherence in the form of broken links wipes out the non-Markovian behaviour and gives rise to a Markovian process. On the other hand, it has also been shown that the non-Markovian behaviour can be enhanced when the walker is subjected to some specific static or dynamic disorder \cite{kumar2018}. In this paper, we analyse non-Markovianity for the quantum walk on the line with a single impurity at the origin. In particular, we examine the relationship between non-Markovianity and  emerging bound states depending on the impurity's phase angle. After solving for the bound states by a transfer matrix approach in section~\ref{sec:qwwspi}, we focus on the effects of these bound states on the system's evolution and investigate the reduced coin dynamics from the point of view of quantum non-Markovianity. We characterize the localisation in position space using the properties of bound states and participation ratio of the probability distribution of quantum walks. We compare the trends in these quantities with those from two non-Markovianity measures of the reduced coin dynamics to reveal the connection between the phenomenon of bound state localisation and non-Markovianity. We point out the differences between the outcomes of the two non-Markovianity measures, which are based on state distinguishability and system-ancilla entanglement, in relation with the emergence of bound states. Our results for the localisation and non-Markovianity in comparison with each other follow in section~\ref{sec:locAndNM}. We present a discussion of our results and conclude in section~\ref{sec:conc}.

\section{\label{sec:qwwspi}Quantum walk with a phase impurity}

In analogy to the classical random walk, a single step $\hat{W}$ for the quantum walk on the line is defined to include two consecutive unitary operations, a coin-toss $\hat{C}$ and a conditional translation $\hat{T}$, which is given in the form $\hat{W}=\hat{T}\hat{C}$ \cite{aharonov1993,kempe2003,venegas-andraca2012}. The time evolution therefore takes place in the discrete bipartite coin-position Hilbert space $H_c \otimes H_p$ spanned by the states $\ket{c,n}$, where $c\in \set{\ua,\da}$ describes the two coin states being the eigenstates of the Pauli operator in z-direction, $\hat{\sigma}_z$, and $n$ is an integer labelling the discrete sites that the walker can be found. Thus, $t$ steps of the walk are realized by applying $\hat{W}$ repeatedly to an arbitrary initial state $\ket{\Psi_0}$ and the total state of the system at step $t$ can be written as
\begin{equation}
\ket{\Psi_t} = W^t \ket{\Psi_0} = \sum_{c,n} a_{c,n}(t) \ket{c,n},
\end{equation}
where $a_{c,n}(t)$ are complex coefficients belonging to the appropriate coin and position states at time $t$. The probability to find the walker at any site $n$ after $t$ steps is accordingly given as $P_n(t)\!=\!\sum_c|a_{c,n}(t)|^2$. The coin operator can preferably be chosen as any unitary operator in $SU(2)$. Here, we will restrict it to the rotations about x-axis by an angle of $2\theta \in [0,4\pi]$, which are written as
\begin{equation}
\hat{C}_\theta = \rme^{-i \theta \hat{\sigma}_x } \otimes \hat{I}_N
\end{equation}
where $\hat{\sigma}_x$ is the Pauli-X operator and $\hat{I}_N=\sum_n |n\rangle\langle n|$ is the identity operator acting on the position space with dimension $N$. As it is seen, the coin operator only affects the coin state and keeps the position state intact. The conditional translation operator on the other hand moves the walker to the left or to the right depending on the coin state shown here by 
\begin{equation}
\hat{T} =
\sum_n 
\bigg[
\outerp{\ua}{\ua} \otimes \outerp{n+1}{n}
+ 
\outerp{\da}{\da} \otimes \outerp{n-1}{n}
\bigg],
\end{equation}
and thus entangling the coin and the position degrees of freedom in general~\cite{carneiro2005}.

\begin{figure}[t]
\centering
\includegraphics[scale=0.75]{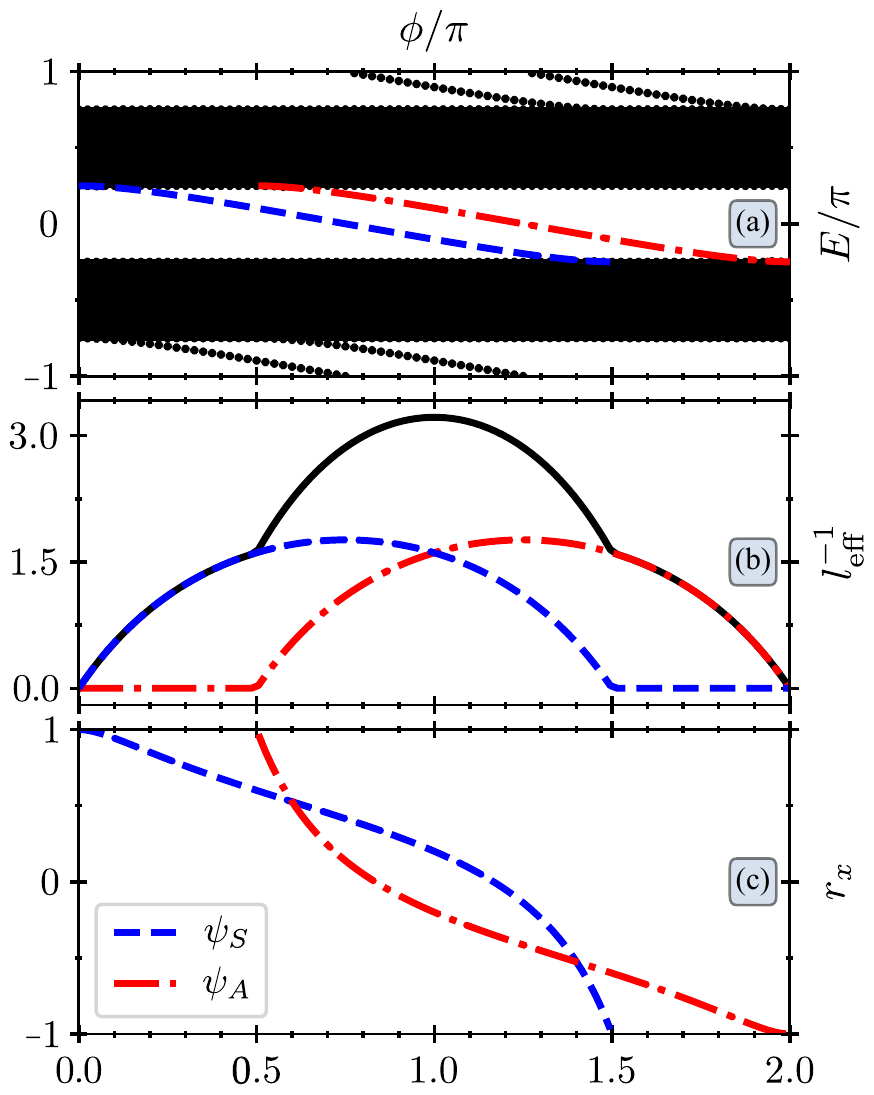}
	\caption{(a) Energy band diagram of the walk ($\theta=\pi/4$) as a function of the phase parameter $\phi$, where $\phi=0$ corresponds to the standard quantum walk. The analytic result given in (\ref{eq:sol1}) is drawn for reflection symmetric (blue dashed) and anti-symmetric (red dot-dashed) bound state quasi-energy values, where the numerical values (black dots) are omitted since they exactly coincide with the analytic result. (b) Inverse localisation length of reflection symmetric (blue dashed) and anti-symmetric (red dot-dashed) energy eigenstates and their sum as the effective inverse localisation length (solid black) as a function of $\phi$, which is used as a measure of localisation. (c) Reduced coin density matrix parameter $r_x$ for reflection symmetric (blue dashed) and anti-symmetric (red dot-dashed) bound states as a function of $\phi$ gives average value about which oscillations in the reduced density matrix take place.
\label{fig:bandStr}}
\end{figure}

The walk can also be considered as a stroboscopic simulation of a quantum evolution generated by an effective Hamiltonian $\hat{H}_\theta$ such that $\hat{W}_\theta\equiv\exp{(-i \hat{H}_\theta)}$, where we assume that the time required for taking one step and $\hbar$ are both set to unity \cite{kitagawa2010,kitagawa2012}. It is well-known that the spectrum of this Hamiltonian has a band structure with a period of $2\pi$, which arises from the discrete time-translation symmetry of the walk, i.e., the Hamiltonian is in fact a representative of a recurring single-step evolution (see figure~\ref{fig:bandStr} (a)). The energy eigenvalues $E$ here are called quasi-energies, similar to the quasi-momentum $k$ showing up due to discrete spatial translation symmetry of $\hat{W}_\theta$ in the standard quantum walk by a unit lattice spacing. The standard walk Hamiltonian becomes diagonal in this quasi-momentum basis via the transformation $\ket{c,k}=\frac{1}{\sqrt{2\pi}}\sum_n \rme^{-ikn}\ket{c,n}$ with $k\in [-\pi,\pi]$ and one obtains the dispersion relation $\cos E(k) = \pm \cos \theta \cos k$, where $k$ is in units of $\hbar$ over the lattice spacing. Two quasi-energy bands associated with the coin states are symmetric about $E=0$ and the band gap closes for $\theta=0$ or $\pi$. Any pair of eigenstates with a quasi-energy difference of $\pi$ can be associated with one another via the sublattice operator 
\begin{equation}
\hat{S} = \hat{I}_2 \otimes \sum_n (-1)^n \outerp{n}{n},
\label{eq:sublatOpr}
\end{equation}
which is both unitary and Hermitian, i.e., $\hat{S}^2=\hat{I}_{2N}$ \cite{asboth2012}. Thus, for each eigenstate $|E\rangle $ of $\hat{H}_\theta $ with quasi-energy $E$ in a given band, there exists another eigenstate $\hat{S}|E\rangle $ with quasi-energy $E+\pi$ in the other band, which can actually be deduced directly using  (\ref{eq:sublatOpr}),
\begin{equation}
\hat{S} \hat{W}_\theta \hat{S} = -\hat{W}_\theta~~
\Rightarrow~~
\hat{S} \hat{H}_\theta \hat{S} = \hat{H}_\theta + \pi.
\label{eq:enAlt}
\end{equation}
The sublattice operator has two degenerate eigenvalues $\pm 1$ such that eigenstates corresponding $+1$ ($-1$) parity, which we will denote by $\ket{S_e}$ ($\ket{S_o}$), occupy only even (odd) labelled sites in the position space. These eigenstates can also be written in terms of the symmetric (for~$+1$) and anti-symmetric (for~$-1$) superpositions of $\ket{E}$ and $\hat{S}\ket{E}$ such that
\begin{equation}
\ket{S_{e,o}} = \ket{E} \pm \hat{S} \ket{E}.
\end{equation}
Note that the step operator $\hat{W}_\theta$ transforms $\ket{S_{e}}$ to $\ket{S_{o}}$, or vice versa, up to an overall phase of $\rme^{-iE}$. Therefore, if one initializes the walk with either of $\ket{\Psi_0}=\ket{S_{e,o}}$, the total state will oscillate from one to the other forever, i.e., step operator moves the state back and forth between two sublattices.  It is worth mentioning here that this periodic oscillation with a period of $2$ will play an important role in our discussion of non-Markovianity. 

We will study the non-Markovian behaviour from the point of view of the coin sub-system, which will be considered as an open system with the position space interacting with it as the environment \cite{hinarejos2014}. Therefore, the time evolution of the coin density matrix will be investigated. Since the stationary states of the standard quantum walk are product states of the form $\ket{E_k} = \ket{\chi_k}\otimes\ket{k}$, the reduced pure coin density matrix can be written as $\rho^\mathrm{coin}_k=\outerp{\chi_k}{\chi_k} = (I_2 + \vec{r}_k \cdot \sigma)/2$, where
\begin{equation}
\vec{r}_k = \frac{1}{\sin E(k)} (\cos k \sin \theta, - \sin k \sin \theta, -\sin k \cos \theta).
\label{eq:ideal_rxyz}
\end{equation}

\begin{figure}[t]
\centering
\includegraphics[scale=1.55]{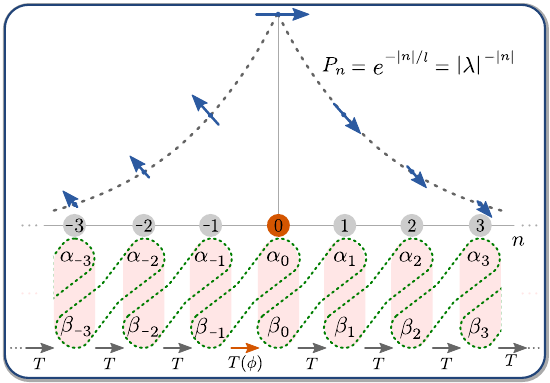}
\caption{Schematic representation of a bound state and its probability distribution over the position space. The walker acquires a phase $\rme^{i\phi}$ due to the impurity at the origin.  As given in (\ref{eq:trRec}), the coefficients $(\alpha_n,\beta_{n-1})$ are related to each other by transfer matrices $T$ and $T(\phi)$ for $n\neq 0$ and $n=0$, respectively. The localisation length $l$ is determined from eigenvalue $\lambda$ of the transfer matrix $T$. Blue arrows represents the coin state in the phase space and their lengths are drawn proportional to $P_n$.
\label{fig:transfer}}
\end{figure}

We consider a modified version of the standard quantum walk such that whenever the walker passes through the origin $n\!=\!0$, it acquires a phase of $\rme^{i\phi}$ as illustrated in figure~\ref{fig:transfer}. This effect can be introduced to the step operator by rewriting it as $\hat{W}_\theta'=\hat{T} \hat{C}_\theta \hat{P}\equiv \rme^{-i\hat{H}_\theta'}$ to include the phase operator 
\begin{equation}
\hat{P} = \hat{I}_2 \otimes \sum_n \rme^{i \phi_n } \outerp{n}{n},
\label{eq:phase_op}
\end{equation}
where $\phi_n=\phi\delta_{n,0}$ and $\delta_{n,m}$ denotes the Kronecker delta function. This model was studied by Wojcik et al. and the bound eigenstates of the double step operator $\hat{W}_\theta'^2$ are obtained \cite{Wojcik2012}. We will provide an alternative solution for the stationary bound states in this model for the single step operator $\hat{W}_\theta'$ using a transfer matrix approach \cite{zhao2015disordered}. We note that a double step operator would be useless in our case since it restricts the evolution to one of the sublattices, and hence, no oscillation takes place between $\ket{S_{e,o}}$ and $\ket{S_{o,e}}$ contrary to the discussion we will have, where these oscillations lie at the center of non-Markovian behaviour.

The phase operator $\hat{P}$ breaks the translation invariance of the step operator $\hat{W}_\theta'$ in the considered model. However, we can still employ the reflection symmetry which is introduced by the reflection operator
\begin{equation}
\hat{R} = \sigma_x \otimes \sum_n |-n\rangle \langle n |,
\label{eq:reflOpr}
\end{equation}
in finding the stationary states of $\hat{H}_\theta'$ as follows. Similar to the sublattice operator, $\hat{R}$ also has the property $\hat{R}^2=\hat{I}_{2N}$ and it possesses two degenerate eigenvalues $\pm 1$. Also, by considering the commutation relations that $[\hat{R},\hat{H}_\theta']=0=[\hat{R},\hat{S}]$, eigenstates of $\hat{H}_\theta'$ can be labelled by a definite $\pm$ parity, i.e. $\hat{R} | E^{\pm} \rangle =\pm | E^{\pm} \rangle$, and application of $\hat{S}$ does not change this parity and it yields the eigenstates with quasi-energies $E^\pm+\pi$ since the condition (\ref{eq:enAlt}) is still valid for $\hat{W}_\theta'$. These energy eigenstates can be written in the component form
\begin{equation}
\ket{E^\pm} =\sum_{n} \alpha^\pm_n \ket{\uparrow,n}+\beta^\pm_n\ket{\downarrow,n},
\label{eq:stStates}
\end{equation}
where the coefficients obey the constraint $\alpha^{\pm}_{-n} = \pm \beta^{\pm}_{n}$ 
as a direct consequence of the reflection symmetry. Thus, $\ket{E^\pm}$ can be constructed by knowing only half of their components. Also, by using $\hat{W}_\theta'\ket{E^\pm}~=~\exp(-i E^\pm)\ket{E^\pm}$, one can obtain the recursion relations between these coefficients and rearrange them in the following way 
\begin{equation}
\begin{pmatrix}
	\alpha_{n+1}^\pm \\
  \beta_n^\pm
\end{pmatrix}
=
T(\phi_n)
\begin{pmatrix}
    \alpha_{n}^{\pm} \\
    \beta_{n-1}^{\pm} 
  \end{pmatrix}.
\label{eq:trRec}
\end{equation}
Here, the matrix $T(\phi_n)$ is called the transfer matrix for site $n$ and it connects the adjacent coefficients in (\ref{eq:stStates}) (See figure~\ref{fig:transfer}), which is given in general as
\begin{equation}
T(\phi_n)=
\begin{pmatrix}
\rme^{i(E^\pm+\phi_n)}\sec\theta & -i\tan\theta \\
i\tan\theta & \rme^{-i(E^\pm+\phi_n)}\sec\theta 
\end{pmatrix},
\end{equation}
with the inverse $T^{-1}(\phi_n) = \sigma_x T(\phi_n) \sigma_x$.
Reflection symmetry implies $T(\phi_{-n}) = \sigma_x T^{-1}(\phi_n) \sigma_x=T(\phi_{n})$ and thus $\phi_n = \phi_{-n} $. 
Since $\phi_n = \phi \delta_{n,0}$ in our model, we set $T(\phi_n=0) = T$ for $n \neq 0$. Now, the problem of finding the stationary states is boiled down to determine a suitable pair $(\alpha_1^{\pm},\beta_0^{\pm})^T$ in (\ref{eq:stStates}) satisfying the reflection property. We can look for stationary states whose coefficients as simultaneous eigenvectors of the transfer matrices must satisfy
\begin{equation}
T(\phi) \sigma_x \!
\begin{pmatrix}
    \alpha_{1}^{\pm} \\
    \beta_{0}^{\pm} 
  \end{pmatrix} \!\!
  = \!
\pm \!
\begin{pmatrix}
    \alpha_{1}^{\pm} \\
    \beta_{0}^{\pm} 
  \end{pmatrix}
\mathrm{and}~
T \!
\begin{pmatrix}
    \alpha_{n}^{\pm} \\
    \beta_{n-1}^{\pm} 
  \end{pmatrix} \!\!
= \!\!
\lambda \!
\begin{pmatrix}
    \alpha_{n}^{\pm} \\
    \beta_{n-1}^{\pm} 
\end{pmatrix}
\label{eq:eigvalue_eq_tmatrix}
\end{equation}
such that
\begin{equation}
\begin{pmatrix}
    \alpha_{n+1}^{\pm} \\
    \beta_{n}^{\pm} 
  \end{pmatrix}
=
\lambda^n
\begin{pmatrix}
    \alpha_{1}^{\pm} \\
    \beta_{0}^{\pm} 
  \end{pmatrix}, \qquad n\ge 1.
\label{eq:eigvalue_eq_tmatrix2}
\end{equation}
In the infinite chain limit, vanishing boundary conditions impose that $|\lambda|<1$ for bound states and matching the two forms 
\begin{eqnarray}
\begin{pmatrix}
    \alpha_{1}^{\pm}\\
    \beta_{0}^{\pm}
\end{pmatrix}
&=& 
C(E^\pm)
\begin{pmatrix}
    \sin \theta \\
    \sin (E^\pm) - i \sqrt{\sin^2 \theta - \sin^2 (E^\pm) }  
\end{pmatrix}, \nonumber \\
\begin{pmatrix}
    \alpha_{1}^{\pm} \\
    \beta_{0}^{\pm} 
\end{pmatrix}
&=&
C(E^\pm) \sin \theta
\begin{pmatrix}
    \rme^{i(E^\pm+\phi)} \\
    \pm \rme^{\pm i \theta} 
\end{pmatrix}
\label{eq:transfer_eigvecs}
\end{eqnarray}
gives the quasi-energy
\begin{equation}
E^{\pm}  = \cot^{-1} \left( \pm \frac
{1- \sin (\theta \mp \phi) \sin \theta }
{\sin \theta \cos (\theta \mp \phi) }  \right),
\label{eq:sol1}
\end{equation}
provided that
\begin{equation}
\sin {(E^{\pm}+\phi \mp \theta)} = \pm \frac{ \sqrt{\sin^2 \theta  - \sin^2 E^{\pm} }}{\sin \theta} .
\label{eq:sol1_cond}
\end{equation}
The normalization constant is given by $|C(E^{\pm})|^2=\left(1-\lambda (E^{\pm})^2\right)/2$ with the transfer matrix eigenvalue 
\begin{equation}
\lambda_\pm \!\!=\!\!\lambda (E^{\pm})
\!=\! \frac{1}{\cos \theta} \! \left(\! \cos (E^{\pm}) \!-\! \sqrt{\sin^2 \theta \!-\! \sin^2 (E^{\pm})} \!  \right)\!\!.
\label{eq:eigenvalues}
\end{equation}
It is seen that $|\lambda_{\pm}|\le 1$ provided that $|\sin E^\pm| \le |\sin \theta|$.
When a solution with definite parity and an eigenvalue $-\theta < E^\pm < \theta$ exists, then there is a sublattice symmetric solution, which  has $\lambda_\pm \rightarrow -\lambda_\pm $ and 
$\begin{pmatrix}
    \alpha_{1}^{\pm} \\
    \beta_{0}^{\pm} 
\end{pmatrix} \rightarrow
\hat{S}
\begin{pmatrix}
    \alpha_{1}^{\pm} \\
    \beta_{0}^{\pm} 
\end{pmatrix}$, with the same parity and shifted eigenvalue $-\theta+\pi < E^\pm < \theta+\pi$.
The combinations of the sublattice symmetric pairs of reflection symmetric states, for example, give $\ket{S^+_{e,o}}\!=\!\ket{E^+} \pm \ket{E^++\pi}$ which are supported on even/odd sites. Similar combinations $\ket{S^-_{e,o}}$ exist for reflection anti-symmetric bound states.

The reflection symmetric bound states exist only in the interval $\phi \in (0 , 2\pi- 2\theta )$ and the reflection anti-symmetric bound states exist for $\phi \in (2\theta , 2\pi)$. The quasi-energies are shown with blue dashed and red dot-dashed lines in the quasi-energy diagram as a function of $\phi$ in figure~\ref{fig:bandStr}(a) for $\theta=\pi/4$, which corresponds to balanced walks.
The numerical solution is performed for a large lattice with periodic boundary conditions and the results are indistinguishable on this scale from the quasi-energies of the bound states determined from the eigenvalues of the transfer matrix.

We would like to quantify the localisation in our model as a function of the impurity phase $\phi$. For this purpose one of the quantities we calculate is an effective inverse localisation length $\ell_\mathrm{eff}^{-1}$ which is the sum of the inverse localisation lengths over all energy eigenstates. Since the localisation length diverges for an extended state in the infinite chain limit, extended states' contribution to $\ell_\mathrm{eff}^{-1}$ also vanishes in that limit and we are left with only the inverse localisation lengths of the bound states given by the eigenvalues of the transfer matrix, i.e. $\ell^{-1} = \ln \lambda$. Therefore, the sum of $\ell^{-1}$ over all stationary states is determined by the number of  bound states and their localisation lengths. The total $\ell_\mathrm{eff}^{-1}$ (solid black curve) is plotted in figure~\ref{fig:bandStr}(b) as a function of $\phi$, where the individual contributions of symmetric (blue dashed) and anti-symmetric (red dot-dashed) bound states are also shown. The localisation length of a bound state is minimum when its quasienergy is at the center of the band gap (see (\ref{eq:eigenvalues})), so that the reflection symmetric and anti-symmetric bound states become maximally localised when $\phi=3 \pi/4$ and $\phi=5 \pi/4$, respectively. The total $\ell_\mathrm{eff}^{-1}$ is symmetric about $\phi=\pi$ where it attains its maximum value. Although it monotonically decreases towards both sides of $\phi=\pi$, kinks in $\ell_\mathrm{eff}^{-1}$ occur at $\phi=\pi/2$ ($\phi=3 \pi/2$) at the (dis)appearance of new bound states as $\phi$ increases from $0$ to $2\pi$.

Unlike the homogeneous quantum walk, in the presence of phase impurity the energy eigenstates become entangled in the composite coin-position space in this model and $\rho^\mathrm{coin}$ becomes a mixed state. Furthermore, the $y$- and $z$-components of the density matrix vector $\vec{r}$ become zero for all energy eigenstates (see (\ref{eq:ideal_rxyz}) for the standard walk). Here, we note the $x$-components of the reflection symmetric and anti-symmetric bound states
\begin{equation} 
r_{x,\pm} = \pm \frac{1}{2} 
\left[2\lambda_{\pm} \cos \left(E^{\pm}+\phi \mp \frac{\pi}{4}\right)
+\left(1-\lambda_{\pm}^2\right)\right].
\end{equation}
As $\phi$ changes, $r_x$ changes from $+1$ to $-1$ for both the symmetric and anti-symmetric bound states over the regions of their existence as shown in figure~\ref{fig:bandStr}(c). (Since the action of the unitary sublattice operator on the coin space is trivial, sublattice symmetric bound state give the same  $\rho^\mathrm{coin}$.) At $\phi \sim 0.6 \pi $ and $ \phi \sim 1.4 \pi $, the difference in $r_{x,\pm}$ becomes zero, which means the two coin states become indistinguishable for bound states of either parity. 

\section{\label{sec:locAndNM}Localisation \& Non-Markovianity}

\subsection{Bound state localisation}

\begin{figure}[t]
\centering
\includegraphics[scale=0.95]{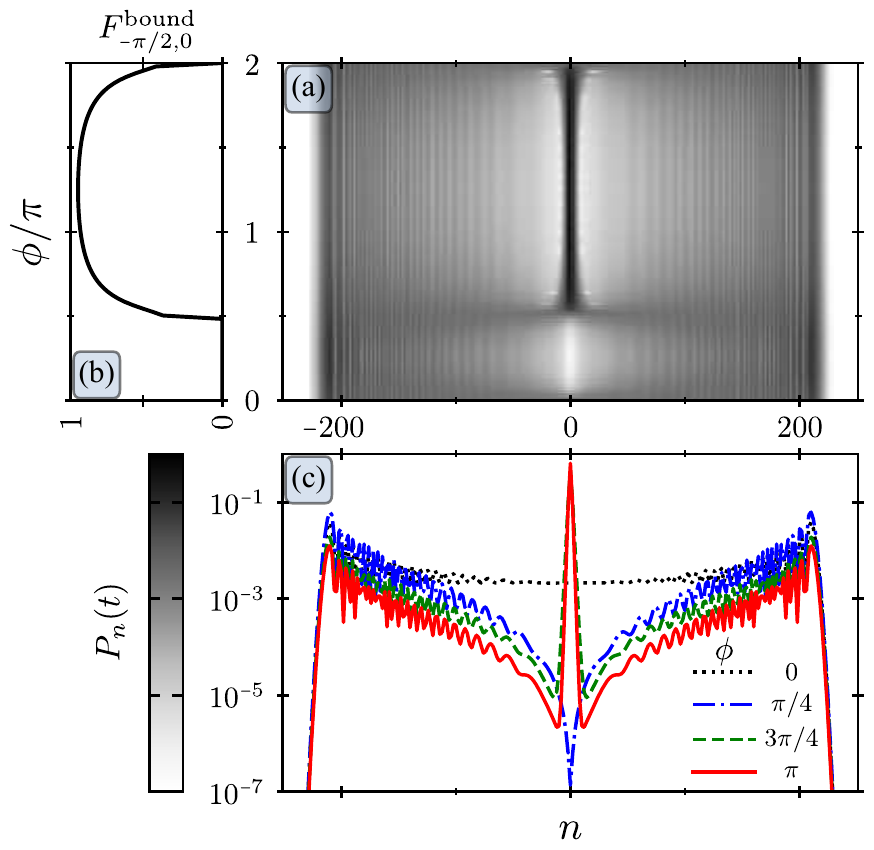}
\caption{(a) The probability distribution $P_n(t)$ of the walk after $300$ steps for the reflection anti-symmetric initial state $\ket{\Psi_\textrm{A}}$ as a function of the impurity phase angle $\phi$. (b-c) Depending on the overlap $F_{-\pi/2,0}^\textrm{bound}$ between $\ket{\Psi_\textrm{A}}$ and bound states, the probability distribution can have a peak or a dip at the starting site of the walker.
\label{fig:single_state_distributions}}
\end{figure}

We commence this section by presenting our results for time evolving states in balanced quantum walks ($\theta=\pi/4$) making connection to the stationary bound states obtained in the previous section. In general, the dynamical properties resulting from the evolution of the quantum walk depend on the initial state. We consider initial states such as $\ket{\Psi_0}=\ket{\psi_c} \otimes \ket{0}$ which are localised at the origin, for examining the effects due to the existence of bound states around the origin. Here, $\ket{\psi_c}$ is the initial coin state and the reflection symmetry of $\ket{\Psi_0}$  is defined by the symmetry of $\ket{\psi_c}$ under rotation by $\pi$ about the $x$-direction. For our purposes, we choose to concentrate on the reflection symmetric and anti-symmetric initial states whose coin subsystems are the eigenstates of $\sigma_x$, i.e., the initial states $\ket{\Psi_\textrm{S}}= \frac{1}{\sqrt{2}} [\ket{\ua}+\ket{\da}]\otimes\ket{0}$ and $\ket{\Psi_\textrm{A}}~=~\frac{1}{\sqrt{2}} [\ket{\ua}-\ket{\da}]\otimes\ket{0}$, respectively. Time evolution starting from these states yields symmetric probability distributions about the origin independent of the presence of a phase impurity. For example, in case of $\ket{\Psi_0}\equiv \ket{\Psi_\textrm{A}}$, figure~\ref{fig:single_state_distributions}(a) shows the probability distribution in position $P_n(t)$ at $t=300$ for all $\phi$ values. The significance of the reflection anti-symmetric bound state for a given $\phi$ becomes clear when we compare the degree of localisation and the overlap of the initial state with the corresponding bound states $\ket{E^-}$. For an arbitrary initial coin state $\ket{\psi_c}=\cos\frac{\gamma}{2}\ket{\ua}+\rme^{i\eta}\sin \frac{\gamma}{2}\ket{\da}$, this overlap is given by 
\begin{eqnarray}
F^\mathrm{bound}_{\gamma,\eta} &=&
\sum_{E \in \{ E_\mathrm{bound} \}}  | \braket{E}{\Psi_t} |^2
\nonumber \\
&=&
\frac{(2-\lambda_+^2-\lambda_-^2)-(\lambda_+^2-\lambda_-^2)\sin\gamma\cos\eta}{2}
\label{eq:pBound}
\end{eqnarray}
For our selection of initial coins, (\ref{eq:pBound}) simplifies to $F^\mathrm{bound}_{\pm\pi/2,0} = 1-\lambda_\pm^2$. The localisation region apparent in figure~\ref{fig:single_state_distributions}(a), when $\phi \in (\pi/2,2\pi)$, is directly related with $F^\mathrm{bound}_{-\pi/2,0}$ shown in figure~\ref{fig:single_state_distributions}(b). The walker is most localised when the overlap becomes maximum at $\phi=5 \pi/4$. Similarly, in the interval $\phi \in (0 , \pi/2 )$ the overlap with the bound subspace is zero and the probability distribution has a dip around the initial site. Four representative probability distributions for different $\phi$ values are shown in detail in figure~\ref{fig:single_state_distributions}(c). For the standard quantum walk $\phi=0$, the probability distribution is uniform in the middle region and there are two peaks near the edges which move in opposite directions with constant speed. For any $\phi$, the resulting probability distribution spreads balistically in the position space, i.e., it's variance is proportional to $t^2$. However, the proportionality constants may vary depending on the amount of localisation, and hence, on $\phi$.

\begin{figure}[!t]
\centering
\includegraphics[scale=0.85]{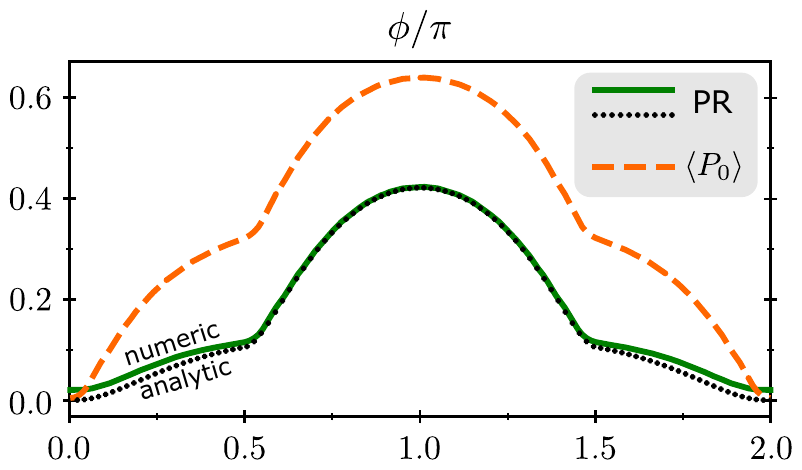}
\caption{The numerical results for the participation ratio (PR) and the average probability at the origin $\left<P_0\right>$ with respect to $\phi$ after $150$ steps. The analytical prediction for PR (black dots) is also provided.
\label{fig:average_loc}}
\end{figure}

The average probability distribution of the walk $\left<P_n(t)\right>$ is obtained by averaging $P_n(t)$ over all possible initial coin states. However, we observe that we get exactly the same result by only taking into account any pair of orthogonal coin states. This is due to the fact that the average probability distribution resulting from two walks starting with any two orthogonal coin states at the origin is equal to the one resulting from the evolution of a completely mixed coin state. (The resulting distribution is symmetric since the completely mixed coin state at the origin is reflection invariant.)  Also, for the long-time limit, the bound states stay in the vicinity of the origin, whereas the extended states get spread over the infinite position space yielding probabilities going to zero. Based on these facts, we can obtain an analytic expression to estimate the long-time behaviour of $\left<P_n(t)\right>$ by projecting the evolved state onto the bound subspace and averaging the corresponding probabilities over two orthogonal initial states, such that
\begin{eqnarray}
\left< P_0 \right > &=& \frac{1}{2}\left[(1-\lambda_+^2)^2 + (1-\lambda_-^2)^2\right]~~\textrm{and} \\
\left< P_n \right >
&=&
\frac{1}{4}
[\lambda_+^{2|n|-2} (1+\lambda_+^2) (1-\lambda_+^2)^2
\nonumber \\
&+&
\lambda_-^{2|n|-2} (1+\lambda_-^2)  (1-\lambda_-^2)^2],
\label{eq:Pn}
\end{eqnarray}
where $n\neq 0$ and non-zero probabilities appear for even (odd) sites only after even (odd) number of steps. To quantify the localisation, we utilize the participation ratio of the averaged probability distribution, which is given by
\begin{equation}
\mathrm{PR} = \sum_n \left<P_n(t)\right>^2.
	\label{eq:PR}
\end{equation}
For a uniform probability distribution over $N$ sites, PR yields its minimum value $\sim N^{-1}$. At the other extreme of localisation at one site, PR takes its maximum value of one. In figure~\ref{fig:average_loc}, the numeric results for the PR (green solid curve) and $\left<P_0\right>$ (orange dashed curve) for $150$ steps are represented. Both of them is calculated by using the average probability distribution $\left<P_n(t)\right>$ which is averaged over a pair of orthogonal initial coin states as we mentioned before. We also provide the analytic prediction of PR (black dots) for the long-time behaviour using (\ref{eq:Pn}) and (\ref{eq:PR}) which slightly differs from its numerical simulation, whereas we omitted that of $\left<P_0\right>$ for clarity since it exactly fits to the numerical data. First of all, both curves exhibit similar behaviour with respect to $\phi$ and $\left<P_0\right>$ pointing out that localisation occurs around the impurity site. They get maximized at $\phi=\pi$ and vanish at the standard quantum walk limit $\phi=0,2\pi$. The kinks at $\phi=\pi/2,3\pi/2$ are due to bound states appearing or disappearing in this model as discussed previously. This behaviour matches exactly that of the effective localisation length determined by the bound states in figure~\ref{fig:bandStr}(b), which consequently shows that the localisation properties of the walk in the long-time limit is determined by the number and character of the stationary bound states. The slight difference between the numerical and analytical results of PR stems from the finite number of time steps in the numerical simulation and the fact that contribution from the extended states is completely excluded in the analytical expression. As a consequence of this, the numerical data stays above the analytical prediction. For example, as we approach the standard walk case, the wavefunction for a finite-step walk stays relatively ``localised'' in comparison to that of the long-time case which spreads infinitely over the position space without any localisation. Hence, the numerical prediction  will become zero in the standard walk in this limit as well. The very good agreement between the numerical and analytical results in figure~\ref{fig:average_loc} implies that the effect of the extended states on the PR is negligible even after $150$ steps.

\subsection{Non-Markovianity}

We now turn our attention to the non-Markovian behaviour of the dynamics of the coin for the quantum walk with a phase impurity. As mentioned before, we are interested in the effects of localised bound states and their symmetry on the degree of non-Markovianity of the reduced coin evolution. In order to quantify the amount of memory effects in the open system dynamics from different perspectives, we will comparatively study two well-established measures of quantum non-Markovianity that are based on the information flow dynamics between the coin and the spatial degrees of freedom.

Let us first briefly discuss how to characterize the non-Markovian nature of an open system evolution and identify the existence of possible memory effects in the dynamics. Assume that we have a quantum map \ALS{$\Lambda_{t,0}$}{$\Lambda(t,0)$}, i.e., a completely positive trace preserving (CPTP) map describing the evolution of the open quantum system. The property of divisibility implies that divisible maps satisfy the decomposition rule $\Lambda(t,0) = \Lambda(t,s) \Lambda(s,0)$, where $\Lambda(t,s)$ is a CPTP map for all $s\leq t$. Markovian or so-called memoryless dynamical maps are recognized as the ones that satisfy this decomposition rule. On the other hand, when the divisibility rule is violated, i.e., when \ALS{$\Lambda_{t,s}$}{$\Lambda(t,s)$} is not a CPTP map or when it does not even exist, then the dynamical map $\Lambda$ is said to be non-divisible and the evolution it describes non-Markovian. The concept of divisibility can also be discussed in the context of discrete dynamics, such as quantum walk, where $t,s \in \mathbb{N}$~\cite{luoma15}.

The first non-Markovianity measure that we utilize in our work is known as \ALS{the BLP}{Breuer-Laine-Piilo (BLP)} measure~\cite{breuer09} which is based on the idea of distinguishability of two open system states under a given dynamical evolution. In this approach, the changes in the distinguishability between two arbitrary initial states of the open system during the dynamics are interpreted as the information flow between the open system and its environment. In particular, if distinguishability between the initial states decreases monotonically in time throughout the evolution, the dynamics is said to be Markovian, since in this case information flows from the open system to its environment in a monotonic fashion. However, if distinguishability temporarily increases during the dynamics, then this is understood as a back-flow of information from the environment to the open system giving rise to non-Markovian memory effects. The distinguishability of two systems can be quantified through trace distance between their density matrices $\rho_1$  and $\rho_2$ as
\begin{equation}
D(\rho_1, \rho_2)\!=\!
\frac{1}{2}
||\rho_1\!-\!\rho_2||_1
\!=\!
\frac{1}{2}
\Tr \left[(\rho_1\!-\!\rho_2)^{\dagger} (\rho_1\!-\!\rho_2)\right]^{1/2}
\label{eq:trace_dist}
\end{equation}
which acquires its maximum value of one, when the states $\rho_1$  and $\rho_2$ are orthogonal. At this point, we should stress that since CPTP maps are contractions for the trace distance, BLP measure vanishes for divisible maps, resulting in a memoryless evolution. However, we also emphasize that it is possible for trace distance to monotonically decrease for certain non-divisible maps as well. Therefore, as is well known in the recent literature, even though widely used as a measure for non-Markovianity on its own, BLP measure is actually a witness for the non-divisibility of quantum dynamical maps. The BLP measure can be expressed in discrete time as \cite{luoma15}
\begin{equation}
{\cal{N}}
=
\max_{\rho_{1,2}}
\sum_{t, \Delta D>0} \Delta D_t
=
\sum_{t} \Delta D_t \Theta(\Delta D_t),
\label{eq:nonmarkov}
\end{equation}
where $\Theta(x)$ denotes the Heaviside step function,
\begin{equation}
\Delta D_t
=
D(\rho_{1,t}, \rho_{2,t})-D(\rho_{1,t-1}, \rho_{2,t-1}).
\end{equation}
and the maximization is carried out over all possible initial state pairs. It has been shown that the pair which maximizes the sum in  (\ref{eq:nonmarkov}) is a pair of orthogonal of states~\cite{wissmann12}. In our analysis, we study the reduced system dynamics of a pair of such initial states, namely, $\ket{\psi_{S,A}}$ introduced before, with opposite reflection symmetry, which will be later on revealed as the optimal initial state pair optimizing the BLP measure.

\begin{figure}[!t]
\centering
\includegraphics[scale=0.90]{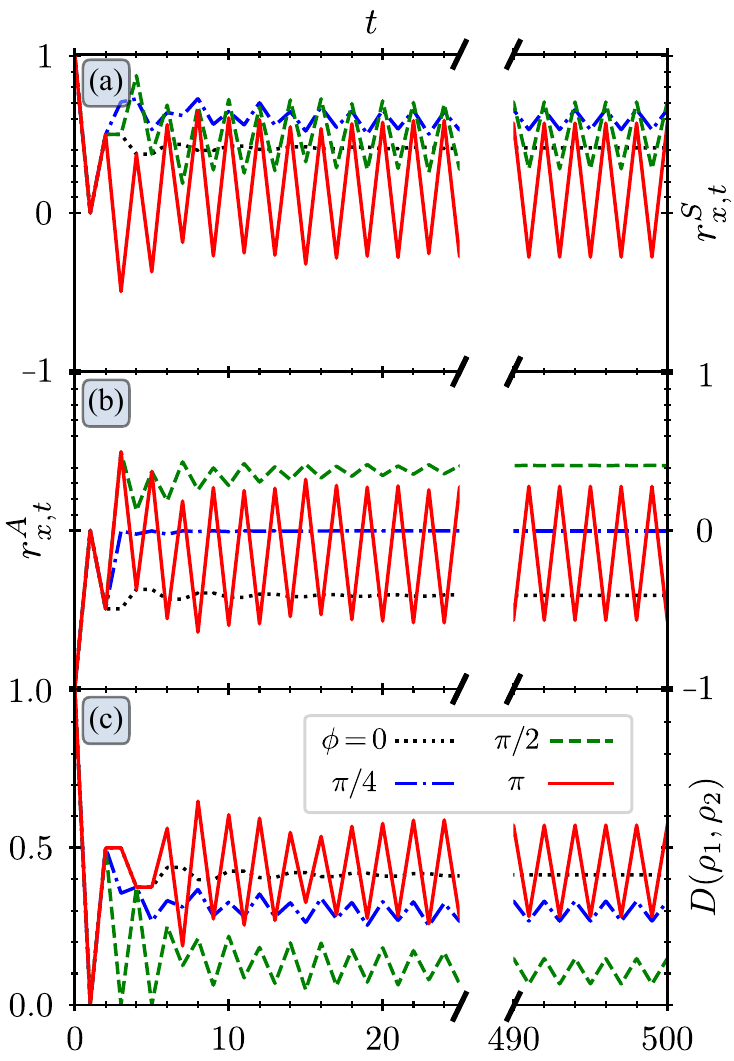}
\caption{Oscillations in the reduced coin density matrices starting from $\ket{\Psi_\text{S}}$ in (a) and from $\ket{\Psi_\text{A}}$ in (b) as a function of time for representative values of the phase parameter $\phi$. The trace distance of these coin states $D(\rho_S,\rho_A)=|r_{x,A}-r_{x,S}|$ is shown in (c) and the oscillating behaviour gives rise to non-zero BLP measure.} 
\label{fig:spinxoscillations}
\end{figure}

The time evolution of $\rho^\mathrm{coin}_{S,A}$ is particularly easy to visualize because the parametrization $\rho^\mathrm{coin}_t = (I + \vec{r}_t\cdot \vec{\sigma})/2$ has only one non-zero component, i.e. $r_{x,t}$, throughout the time evolution which is shown in figure~\ref{fig:spinxoscillations} for representative values of the phase $\phi$. For $\phi=0$, which gives the standart quantum walk, both $r^S_{x,t}$ (black dotted line in figure~\ref{fig:spinxoscillations}(a)) and $r^A_{x,t}=-r^S_{x,t}$ (black dotted line in figure~\ref{fig:spinxoscillations}(b)) undergo damped oscillations with a period of four steps as the  steady-state is reached. Since the oscillations are out of phase for these orthogonal initial states, the trace distance between such states also oscillates in time with decreasing amplitude (black dotted line in figure~\ref{fig:spinxoscillations}(c)). Therefore, even though there is a back-flow of information from the environment to the open system in the standard walk, the damping in oscillations shows that information flow between the two subsystems reduces and eventually vanishes in time~\cite{hinarejos2014}. For non-zero values of $\phi$, oscillations in the initial state component $r^{A(S)}_{x,t}$ arise depending on the overlap with the bound states. When $\phi=\pi/4$, the oscillations in $r^A_{x,t}$ die out very quickly, whereas oscillations with period two between sublattice symmetric pair of localised states survive for $r^S_{x,t}$ as shown by the blue dot-dashed line in figure~\ref{fig:spinxoscillations}(a)-(b). For $\phi=\pi/2$, similar oscillations exist, except they die out more slowly for $r^A_{x,t}$ which has a finite overlap with the emerging reflection anti-symmetric bound state whereas oscillations continue with higher amplitudes for $r^S_{x,t}$ since the reflection symmetric bound-state becomes more localised for this value of $\phi$. At $\phi=\pi$ where bound states of both parities exist, oscillations in $r_{x,t}$ occur with higher amplitudes for both of the initial states in comparison with the other shown phase values.

Having obtained the time dependence of $\rho^\mathrm{coin}_{S,A}$, we calculate the trace distance $D(\rho_S, \rho_A) = |r_{S,x}-r_{A,x}|$, and display our findings in figure~\ref{fig:spinxoscillations}(c), as a function of $\phi$. In contrast to the standard quantum walk where the trace distance oscillations die out in time, we find that they survive for non-zero $\phi$, as at least one of $r^{S,A}_{x,t}$ keeps oscillating in time. However, we should keep in mind that the value of the trace distance also depends on the mean values $\overline{r^{S,A}_{x,t}}$ about which oscillations take place. For example, when $\phi=\pi/2$ we get oscillations in $D(\rho_1,\rho_2)$ with smaller amplitudes than in $r^S_{x,t}$, which will be of importance in our later discussions.

\begin{figure}[t]
\centering
\includegraphics[scale=0.80]{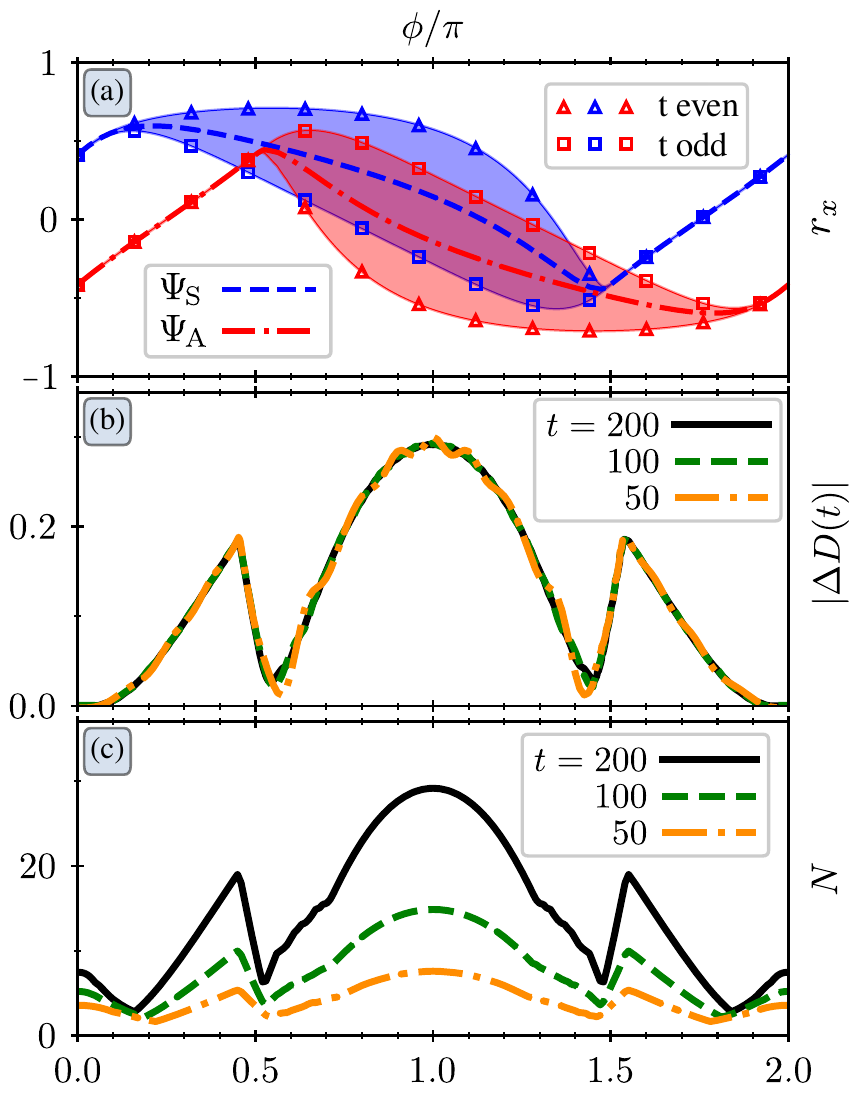}
\caption{(a) Long-time limit time average of the reduced coin density matrix parameter $r_x$ for reflection symmetric ($\ket{\Psi_\text{S}}$) and anti-symmetric ($\ket{\Psi_\text{A}}$) initial states as a function of $\phi$. (Time average is taken over 100 steps between $t=400$ and  $t=500$.) Instantaneous values at even and odd time steps are shown by square and triangle markers, respectively. (b) Trace distance oscillation amplitudes between initial states $\ket{\Psi_\text{S}}$ and $\ket{\Psi_\text{A}}$ at different times show that they quickly converge to their long-time limit values for all $\phi$. (c) BLP measure $\mathcal{N}$ (\ref{eq:nonmarkov}) at three different times. The maximization is performed over all the initial coin states for quantum walks starting at the impurity site. The linear increase in time reflects trace distance oscillations with constant amplitude. (See (b).) 
\label{fig:oscillations}}
\end{figure}

As the persistent oscillations in trace distance play a crucial role for the evaluation of the BLP measure in our model, the oscillation means $\overline{r^{S,A}_{x,t}}$ and the oscillation amplitudes are plotted in figure~\ref{fig:oscillations}(a). Comparison with figure~\ref{fig:bandStr}(c) reveals that, as the overlap between one of the the initial states and the bound states increases, $\overline{r^{S,A}_{x,t}}$ converges to the $r_x$ of the corresponding bound state and oscillations appear. For the interval $\phi \in (\pi /2, 3 \pi /2)$, $\overline{r^{S,A}_{x,t}}$ becomes the same as $r_x$ in the long time limit. The difference in $\overline{r^{S,A}_{x,t}}$ approaches to zero at $\phi \sim 0.6 \pi$ and $\phi \sim 1.4 \pi$, yielding very small values for the trace distance together with the fact that essentially one of $r^{S,A}_{x,t}$ oscillates about their common mean. For other values of $\phi$, the trace distance is mainly determined by the oscillations in $r^{S,A}_{x,t}$. Since the period of the oscillations is two time steps due to the sublattice symmetry, the changes in trace distance can be obtained by subtracting the value at even time step from the neighbouring odd time step which is plotted in figure~\ref{fig:oscillations} (b) at three different times. These plots clearly demonstrate that the trace distance oscillations quickly converge to their long time limit. As the bound states get more localised for certain $\phi$ values and also the overlap of the initial states with them increases, so do the amplitude of the oscillations in the trace distance.

To evaluate the BLP measure, we maximize the sum of the positive increases in trace distance over all possible orthogonal pairs of initial states starting at the impurity site which is shown in figure~\ref{fig:oscillations}(c) as a function of $\phi$ for three increasing values of time. The result reveals that the pair $\ket{\psi_{S,A}}$ that we used for the preceeding analysis actually maximizes the sum in the BLP measure in the long-time limit. In contrast to the standard walk, the initial states maximizing BLP measure are equal superposition of symmetric and anti-symmetric states and these states do not change under other decoherence mechanisms~\cite{hinarejos2014}. 
Near $\phi=0,\pi/2, 3\pi/2, 2\pi$, where bound states are weakly localised, we find that other orthogonal pairs actually maximize the BLP measure. However these regions get smaller as we consider longer time evolutions. The sudden drop in BLP at $\phi=\pi/2,3\pi/2$ is related to the fact that oscillations take place about similar mean values. More importantly, we establish that the BLP measure of non-Markovianity increases with the emergence of bound states and reaches its maximum value at $\phi=\pi$ when the number and localisation of bound states assumes their maximum, as demonstrated by the effective localisation length in figure~\ref{fig:bandStr}(b). The relation of non-Markovianity and localisation is also apparent comparing the BLP curve with the average PR shown in figure~\ref{fig:average_loc}, which is maximum at $\phi=\pi$.

Next, we consider \ALS{the RHP}{Rivas-Huelga-Plenio (RHP)}~\cite{rivas10} measure of non-Markovianity, which is based on the dynamics of entanglement between the system of interest and an ancillary system. The ancillary system $A$ is assumed to have no dynamics of its own and is completely isolated so that any initial entanglement between the system and the ancilla can be affected by the open system dynamics only. In fact, similar to the BLP measure, this measure is also a witness for the violation of the divisibility. Considering the fact that no entanglement measure $E$ can increase under local CPTP maps, it is rather straightforward to observe that 
\begin{equation}
	E[(\Lambda(t,0) \otimes I) \rho_{\mathrm{coin},A}] \leq E[(\Lambda(s,0) \otimes I) \rho_{\mathrm{coin},A}]
\end{equation}
for all times $0\leq s \leq t$. Hence, any increase in the entanglement between the open system and its ancillary can be understood as a signature of non-Markovian memory effects in the time evolution. In other words, while the entanglement contained in $\rho_{\mathrm{coin},A}$ decreases monotonically for all Markovian processes, non-Markovian behaviour in the dynamics can be captured through the temporary increase of entanglement. In the same spirit of the BLP measure, one can then measure the degree of non-Markovianity using the following quantity:
\begin{equation}
{\cal{I}}^{(E)}
=
\max_{\rho_{CA}}
\sum_{t,\Delta E_\mathrm{CA}>0}
\Delta E_{\mathrm{CA},t}
\end{equation}
where $E_\mathrm{CA}$ denotes the entanglement between the coin and a two level ancillary system. For any entanglement measure $E_{CA}$, the RHP measure is found by maximizing ${\cal{I}}^{(E)}$ over all initial reduced density matrices $\rho_{CA}$ of the composite coin-ancilla system. In order to calculate this measure, we start the evolution from composite initial state  $|\Phi^+ \rangle \vert 0 \rangle = \frac{1}{\sqrt{2}}(|\leftarrow \rangle_C|\downarrow\rangle_A+|\rightarrow \rangle_C |\uparrow \rangle_A)\vert 0\rangle$ and use concurrence~\cite{wooters97} as the entanglement measure. It has been shown that when concurrence is used as entanglement measure, the optimum initial state maximizing the RHP measure is a Bell state, for a single qubit interacting with an environment~\cite{neto16}. 

\begin{figure}[t]
\centering
\includegraphics[scale=0.8]{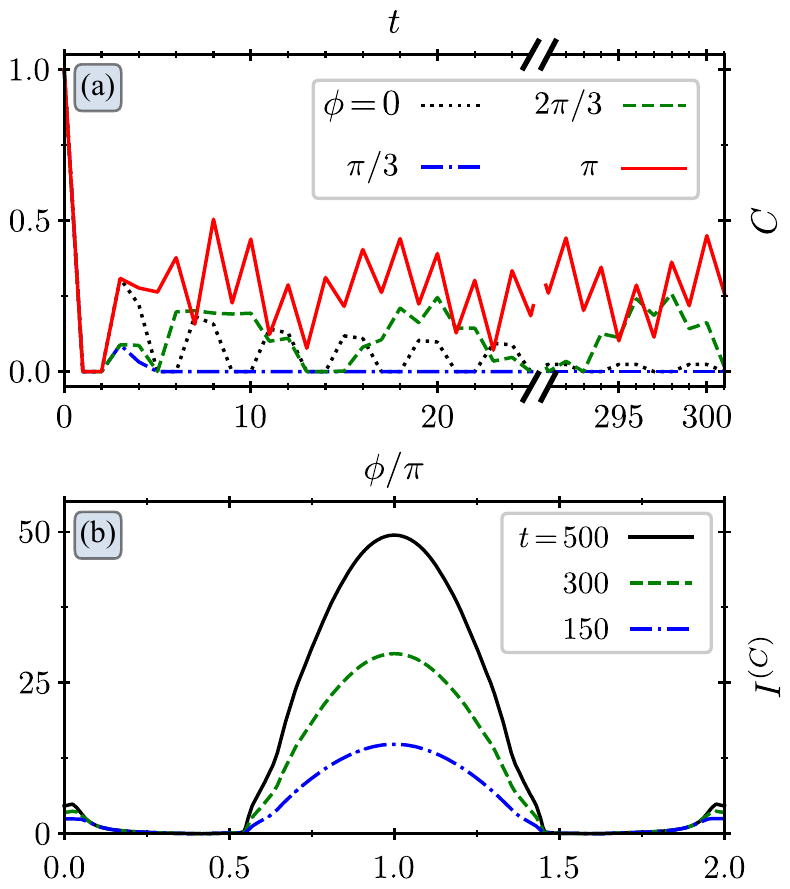}
\caption{(a) Concurrence between the coin and the ancilla qubit as a function of time for representative values of the phase parameter $\phi$. When bound states with both positive and negative reflection parity exist, the concurrence shows oscillations. (See text for the involved frequencies.) (b) Concurrence based RHP measure as a function of $\phi$ at three different time steps showing linear increase with time for $\phi \in (\pi/2,3\pi/2)$. RHP has a vanishing value when well-formed bound states of only positive or negative reflection parity exist.
\label{fig:non_makovianity}}
\end{figure}

Figure~\ref{fig:non_makovianity}(a) shows the variation of the concurrence in time which is calculated from the reduced coin-ancilla state after tracing out the spatial degrees of the walker during the evolution. For the standard quantum walk, the entanglement oscillations with period of four steps are damped and slowly die out with time. Therefore, the RHP measure accumulates a finite amount of non-Markovianity in the long time limit which is similar to the behaviour of the BLP measure for the standard walk. On the other hand, in contrast to the BLP measure, the nature of bound states emerging with non-zero phase $\phi$ plays a key role for the coin-ancilla entanglement. In the presence of reflection symmetric or anti-symmetric bound states only, the concurrence dies out very quickly. This is due to the fact that the symmetric and anti-symmetric states couple to different environmental degrees of freedom. For example, with only symmetric bound states present, the symmetric part of the coin-position state remains mostly localised in the vicinity of the impurity site whereas the anti-symmetric part moves away from the origin. Hence, the coin-ancilla entanglement is quickly destroyed upon tracing out the environmental degrees of position, as the coin-ancilla state becomes an incoherent mixture. An example of this situation is displayed in figure~\ref{fig:non_makovianity}(a) for $\phi=\pi/3$. It is only when both reflection symmetric and anti-symmetric stationary states exist that some entanglement can survive which shows non-decaying oscillations. These oscillations are due to the finite dimension of the bound state subspace and the frequencies of concurrence oscillations can easily be obtained from the quasi-energy differences. Such a case is displayed in figure~\ref{fig:non_makovianity}(a) for $\phi=\pi$ with two dominant periods. One period is of two steps due to the sublattice symmetric bound states with quasi-energy difference $\pi$ and another one is approximately ten steps due to the quasi-energy difference of $\Delta E \approx 0.205\pi$ between reflection symmetric and anti-symmetric states. The latter dependence again shows the importance of bound states of both parities for the RHP measure. The energy difference $\Delta E$ does not change much as $\phi$ changes in the domain of four bound states unless one group of bound states is very weakly bound. (See figure~\ref{fig:bandStr})

Using the time evolution of the coin-ancilla entanglement as shown in figure~\ref{fig:non_makovianity}(a), we evaluate the RHP measure for all values of the impurity phase $\phi$. The results are plotted in figure~\ref{fig:non_makovianity}(b) for three increasing values of the final time. The amount of non-Markovianity measured by the RHP measure drastically depends on whether the reflection symmetric and anti-symmetric bound states are both supported for a given $\phi$ or not. In the interval $\phi \in (0, \pi/2)$ where only the symmetric bound states exist, the concurrence vanishes quickly in time since the coin-ancilla Bell state can only be supported if both symmetric and anti-symmetric bound states exist. Therefore, the coupling of the symmetric and anti-symmetric coin states to different environmental degrees of freedom completely destroys the Bell state of the coin-ancilla system and results in a vanishing value for the RHP measure. A similar situation occurs in the interval $\phi \in (3\pi/2, 2\pi)$ where only reflection anti-symmetric bound states exist and coin-ancilla entanglement is destroyed. In the interval $\phi \in (\pi/2, 3\pi/2)$ where bound states of both symmetries exist, the coin-ancilla entanglement is more robust  and the RHP measure captures the non-Markovianity increasing linearly with $t$ in the long time limit due to non-decaying oscillations in the coin-ancilla entanglement. In this $\phi$ interval, the RHP displays the same behaviour as seen for the BLP measure in figure~\ref{fig:oscillations}(c). 

\section{\label{sec:conc}Conclusion}

We have provided a comprehensive and systematic analysis of non-Markovianity in a quantum walk model with a phase impurity in relation with the phenomenon of localisation. At the heart of analysis lies the manifestation of bound states emerging due to the existence of the phase impurity at the starting site of the walker. We have first presented a technique to analytically obtain the bound states of the model making use of the transfer matrix method. These bound states emerge in one or two sublattice symmetric pairs possessing definite reflection symmetry. With this knowledge at hand, we have explored the localisation properties of the walker in the position space. To this end, we have adopted two initial state independent quantities to measure the degree of localisation, namely, the effective localisation length for all eigenstates and an average participation ratio after time evolution over all initial states starting at the impurity site. Our analysis clearly demonstrates that the degree of localisation of the walker is directly determined by the properties of the bound states.

More importantly, our main contribution in this work is the unveiling of an intrinsic relation between the emergence of bound states and the degree of non-Markovianity of the dynamics of the walker. In order to study non-Markovian behaviour in the time evolution of the walker, after tracing out the spatial degrees of freedom, we have utilized two distinct measures of quantum non-Markovianity, i.e., the BLP and the RHP measures based on the dynamics of trace distance and entanglement, respectively. These measures help us to understand the information flow between the principal coin system and the position system forming the environment from different perspectives. We show that, in the case of the existence of spatial decoherence in the form of a phase impurity, the BLP measure is optimized by the eigenstates of the coin operator for almost all values of the phase $\phi$. Note that when one has decoherence in terms of broken links instead, the degree of decoherence does not change the optimal state maximizing the BLP measure~\cite{hinarejos2014}. Our investigation also proves that phase impurity amplifies the degree of non-Markovianity quantified by the BLP measure.\ALS{, similar to the disorder model studied in~\cite{kumar2018}}{}
The underlying reason behind this behaviour is the oscillations in the state of the coin which essentially takes place between the sublattice symmetric bound state components with a period of two steps. Then, in general, increasing overlap between the initial and the bound states implies a greater degree of non-Markovianity. However, also note that when the time average of the reduced coin states corresponding to two orthogonal initial states are close to each other, the BLP measure drops abruptly.

Next, we employed the RHP measure to analyse the degree of non-Markovianity in the dynamics of the walker. When the coin state is maximally entangled with an ancillary system initially, the amount of entanglement is known to oscillate in time for the standard walk. However, our examination demonstrates that, in case of the existence of a phase impurity, if the bound subspace supports only one type of reflection symmetric state, the coin-ancilla entanglement vanishes after a few time steps and the RHP measure becomes very small compared to the standard walk case. On the other hand, when both reflection symmetric and anti-symmetric bound states are present, the entanglement oscillations are persistent in time, leading to high values of RHP measure. Thus, while the RHP measure is generally in good agreement with the BLP measure when both even and odd parity bound states exist, the RHP measure fails to reliably detect the non-Markovian behaviour when only symmetric or anti-symmetric bound states are present. Most importantly, as can be clearly seen from both measures, maximum non-Markovianity is reached where our localisation measures determined by the bound states become also maximum.
\ALS{}{Relationship between non-Markovianity and localisation have been discussed in random static disorder models~\cite{lorenzo2017quantum,kumar2018} where non-Markovianity increases with disorder.
We observe more nuanced behaviour between bound states and non-Markovianity as discussed above.} 

We would like to indicate that the experimental realization of the model we presented here is quite feasible with today's technology. The time-multiplexing quantum walk employs laser light pulses going successively around a fiber loop where the position space is effectively encoded in the time domain from the point of view of the detectors \cite{schreiber2010}. The main advantage of this setup is it's scalability and it's long coherence times, i.e., it only requires a fixed number of optical elements to realize the quantum walk for relatively large number of steps. The recent developments in the setup allow deterministic out-coupling of the light pulses from any site by utilizing electro-optic modulators \cite{nitsche2018}. It is also possible to introduce arbitrary phases specific to any site by programming of the electro-optic modulators accordingly, which actually would allow the realization of the model we provided here \cite{schreiber2011, nitsche2016}. 

As a concluding remark, it would be interesting to study whether the oscillations due to the bound states become robust in the case of many-body interactions with more degrees of freedom in the context of quantum walks as a future work.

\ack{
\.{I}.Y. is supported by M\v{S}MT under Grant No. RVO 14000 and the Czech Science Foundation under Grant No. GA CR 19-15744Y.
G.K. is supported by the BAGEP Award of the Science
Academy and the TUBA-GEBIP Award of the Turkish Academy of Sciences. G.K. is also supported by the Scientific and Technological Research Council of Turkey (TUBITAK) under
Grant No. 117F317.
B.D. and A.L.S. are supported by Istanbul Technical University Scientific Research Projects Department (ITU BAP No. 40881). A.L.S. would like to acknowledge useful discussions with {\c S}.E. Kocaba{\c s} at earlier stages of this work.
}

\section*{References}
\bibliography{bibliography}

\providecommand{\newblock}{}
\begin{thebibliography}{10}
\expandafter\ifx\csname url\endcsname\relax
  \def\url#1{{\tt #1}}\fi
\expandafter\ifx\csname urlprefix\endcsname\relax\def\urlprefix{URL }\fi
\providecommand{\eprint}[2][]{\url{#2}}

\bibitem{openbook}
Breuer H~P and Petruccione F 2002 {\em The Theory of Open Quantum Systems\/}
  (Oxford University Press)

\bibitem{rivas14}
Rivas A, Huelga S~F and Plenio M~B 2014 {\em Rep. Prog. Phys.\/} {\bf 77}
  094001

\bibitem{breuer16}
Breuer H~P, Laine E~M, Piilo J and Vacchini B 2016 {\em Rev. Mod. Phys.\/} {\bf
  88}(2) 021002

\bibitem{fanchini14}
Fanchini F~F, Karpat G, \ifmmode~\mbox{\c{C}}\else \c{C}\fi{}akmak B, Castelano
  L~K, Aguilar G~H, Far\'{\i}as O~J, Walborn S~P, Ribeiro P~H~S and de~Oliveira
  M~C 2014 {\em Phys. Rev. Lett.\/} {\bf 112}(21) 210402

\bibitem{lu10}
Lu X~M, Wang X and Sun C~P 2010 {\em Phys. Rev. A\/} {\bf 82}(4) 042103

\bibitem{luo12}
Luo S, Fu S and Song H 2012 {\em Phys. Rev. A\/} {\bf 86}(4) 044101

\bibitem{rivas10}
Rivas A, Huelga S~F and Plenio M~B 2010 {\em Phys. Rev. Lett.\/} {\bf 105}(5)
  050403

\bibitem{breuer09}
Breuer H~P, Laine E~M and Piilo J 2009 {\em Phys. Rev. Lett.\/} {\bf 103}(21)
  210401

\bibitem{aharonov1993}
Aharonov Y, Davidovich L and Zagury N 1993 {\em Phys. Rev. A\/} {\bf 48}(2)
  1687--1690

\bibitem{kempe2003}
Kempe J 2003 {\em Contemp. Phys.\/} {\bf 44} 307--327

\bibitem{venegas-andraca2012}
Venegas-Andraca S~E 2012 {\em Quantum Inf. Process.\/} {\bf 11} 1015--1106 ISSN
  1573-1332

\bibitem{ambainis2003}
Ambainis A 2003 {\em Int. J. Quantum Inf.\/} {\bf 01} 507--518

\bibitem{lovett2010}
Lovett N~B, Cooper S, Everitt M, Trevers M and Kendon V 2010 {\em Phys. Rev.
  A\/} {\bf 81}(4) 042330

\bibitem{buerschaper2004}
Buerschaper O and Burnett K  (\textit{Preprint}
  \eprint{arXiv:quant-ph/0406039v2})

\bibitem{oka2005}
Oka T, Konno N, Arita R and Aoki H 2005 {\em Phys. Rev. Lett.\/} {\bf 94}(10)
  100602

\bibitem{kurzynski2011}
Kurzy\ifmmode~\acute{n}\else \'{n}\fi{}ski P and W\'ojcik A 2011 {\em Phys.
  Rev. A\/} {\bf 83}(6) 062315

\bibitem{zhan2014}
Zhan X, Qin H, Bian Z~h, Li J and Xue P 2014 {\em Phys. Rev. A\/} {\bf 90}(1)
  012331

\bibitem{yalcinkaya2015}
Yal\c{c}{\i}nkaya {\.I} and Gedik Z 2015 {\em J. Phys. A\/} {\bf 48} 225302

\bibitem{stefanak2016}
\ifmmode \check{S}\else \v{S}\fi{}tefa\ifmmode~\check{n}\else \v{n}\fi{}\'ak M
  and Skoup\'y S 2016 {\em Phys. Rev. A\/} {\bf 94}(2) 022301

\bibitem{stefanak2017}
{\v{S}}tefa{\v{n}}{\'a}k M and Skoup{\'y} S 2017 {\em Quantum Inf. Process.\/}
  {\bf 16} 72 ISSN 1573-1332

\bibitem{kitagawa2010}
Kitagawa T, Rudner M~S, Berg E and Demler E 2010 {\em Phys. Rev. A\/} {\bf
  82}(3) 033429

\bibitem{kitagawa2012}
Kitagawa T 2012 {\em Quantum Inf. Process.\/} {\bf 11} 1107--1148 ISSN
  1573-1332

\bibitem{Cedzich2016}
Cedzich C, Gr{\"u}nbaum F~A, Stahl C, Vel{\'{a}}zquez L, Werner A~H and Werner
  R~F 2016 {\em J. Phys A: Math. Theor.\/} {\bf 49} 21LT01

\bibitem{Cedzich2018a}
Cedzich C, Geib T, Gr{\"u}nbaum F~A, Stahl C, Vel{\'a}zquez L, Werner A~H and
  Werner R~F 2018 {\em Ann. Henri Poincar{\'e}\/} {\bf 19} 325--383

\bibitem{Cedzich2018}
Cedzich C, Geib T, Stahl C, Vel{\'{a}}zquez L, Werner A~H and Werner R~F 2018
  {\em {Quantum}\/} {\bf 2} 95

\bibitem{brun2003}
Brun T~A, Carteret H~A and Ambainis A 2003 {\em Phys. Rev. A\/} {\bf 67}(3)
  032304

\bibitem{romanelli2005}
Romanelli A, Siri R, Abal G, Auyuanet A and Donangelo R 2005 {\em Physica A\/}
  {\bf 347} 137 -- 152 ISSN 0378-4371

\bibitem{kendon2007}
Kendon V 2007 {\em Math. Struct. Comp. Sci.\/} {\bf 17} 1169--1220

\bibitem{annabestani2010}
Annabestani M, Akhtarshenas S~J and Abolhassani M~R 2010 {\em Phys. Rev. A\/}
  {\bf 81}(3) 032321

\bibitem{schreiber2011}
Schreiber A, Cassemiro K~N, Poto\ifmmode~\check{c}\else \v{c}\fi{}ek V,
  G\'abris A, Jex I and Silberhorn C 2011 {\em Phys. Rev. Lett.\/} {\bf
  106}(18) 180403

\bibitem{karski2009}
Karski M, F{\"o}rster L, Choi J~M, Steffen A, Alt W, Meschede D and Widera A
  2009 {\em Science\/} {\bf 325} 174--177 ISSN 0036-8075

\bibitem{schmitz2009}
Schmitz H, Matjeschk R, Schneider C, Glueckert J, Enderlein M, Huber T and
  Schaetz T 2009 {\em Phys. Rev. Lett.\/} {\bf 103}(9) 090504

\bibitem{zahringer2010}
Z\"ahringer F, Kirchmair G, Gerritsma R, Solano E, Blatt R and Roos C~F 2010
  {\em Phys. Rev. Lett.\/} {\bf 104}(10) 100503

\bibitem{schreiber2010}
Schreiber A, Cassemiro K~N, Poto\ifmmode~\check{c}\else \v{c}\fi{}ek V,
  G\'abris A, Mosley P~J, Andersson E, Jex I and Silberhorn C 2010 {\em Phys.
  Rev. Lett.\/} {\bf 104}(5) 050502

\bibitem{perets2008}
Perets H~B, Lahini Y, Pozzi F, Sorel M, Morandotti R and Silberberg Y 2008 {\em
  Phys. Rev. Lett.\/} {\bf 100}(17) 170506

\bibitem{lorenzo2017quantum}
Lorenzo S, Lombardo F, Ciccarello F and Palma G~M 2017 {\em Scientific
  reports\/} {\bf 7} 42729

\bibitem{cosco2018memory}
Cosco F and Maniscalco S 2018 {\em Physical Review A\/} {\bf 98} 053608

\bibitem{cosco2018bose}
Cosco F, Borrelli M, Mendoza-Arenas J~J, Plastina F, Jaksch D and Maniscalco S
  2018 {\em Physical Review A\/} {\bf 97} 040101

\bibitem{hinarejos2014}
Hinarejos M, Di~Franco C, Romanelli A and P\'erez A 2014 {\em Phys. Rev. A\/}
  {\bf 89}(5) 052330

\bibitem{kumar2018}
Kumar N~P, Banerjee S and Chandrashekar C 2018 {\em Sci. Rep.\/} {\bf 8} 8801

\bibitem{carneiro2005}
Carneiro I, Loo M, Xu X, Girerd M, Kendon V and Knight P~L 2005 {\em New J.
  Phys.\/} {\bf 7} 156

\bibitem{asboth2012}
Asb{\'o}th J~K 2012 {\em Phys. Rev. B\/} {\bf 86} 195414

\bibitem{Wojcik2012}
W{\'{o}}jcik A, {\L}uczak T, Kurzy{\'{n}}ski P, Grudka A, Gdala T and
  Bednarska-Bzd{\c{e}}ga M 2012 {\em Phys. Rev. A\/} {\bf 85} 012329

\bibitem{zhao2015disordered}
Zhao Q and Gong J 2015 {\em Phys. Rev. B\/} {\bf 92} 214205

\bibitem{luoma15}
Luoma K and Piilo J 2016 {\em J. Phys. B\/} {\bf 49} 125501

\bibitem{wissmann12}
Wi\ss{}mann S, Karlsson A, Laine E~M, Piilo J and Breuer H~P 2012 {\em Phys.
  Rev. A\/} {\bf 86}(6) 062108

\bibitem{wooters97}
Wootters W~K 1998 {\em Phys. Rev. Lett.\/} {\bf 80}(10) 2245--2248

\bibitem{neto16}
Neto A~C, Karpat G and Fanchini F~F 2016 {\em Phys. Rev. A\/} {\bf 94}(3)
  032105

\bibitem{nitsche2018}
Nitsche T, Barkhofen S, Kruse R, Sansoni L, {\v S}tefa{\v n}{\'a}k M,
  G{\'a}bris A, Poto{\v c}ek V, Kiss T, Jex I and Silberhorn C 2018 {\em Sci.
  Adv.\/} {\bf 4} 6

\bibitem{nitsche2016}
Nitsche T, Elster F, Novotn{\'y} J, G{\'a}bris A, Jex I, Barkhofen S and
  Silberhorn C 2016 {\em New J. Phys.\/} {\bf 18} 063017

\end{thebibliography}

\end{document}